\documentstyle[12pt,a4,epsf]{article}

\def\mylimit#1{\mathrel{\mathop{\kern0pt\longrightarrow}\limits_{#1}}}

\makeatletter
\@addtoreset{equation}{section}
\makeatother

\newcommand{\VEV}[1]{\left\langle #1 \right\rangle}

\newcommand{\nn}{\nonumber}

\newcommand{\bequ}{\begin{equation}}
\newcommand{\eequ}{\end{equation}}
\newcommand{\beqn}{\begin{eqnarray}}
\newcommand{\eeqn}{\end{eqnarray}}
\newcommand{\bctr}{\begin{center}}
\newcommand{\ectr}{\end{center}}

\begin{document}
\begin{titlepage}

\begin{flushright}
hep-ph/0304076\\
KUNS-1833\\
\today
\end{flushright}

\vspace{4ex}

\begin{center}
{\large \bf
Grand Unification with Anomalous $U(1)$ Symmetry and Non-abelian
Horizontal Symmetry
}

\vspace{6ex}

\renewcommand{\thefootnote}{\alph{footnote}}

Nobuhiro Maekawa\footnote
{e-mail: maekawa@gauge.scphys.kyoto-u.ac.jp
}

\vspace{4ex}
{\it Department of Physics, Kyoto University,\\
     Kyoto 606-8502, Japan}\\
\end{center}

\renewcommand{\thefootnote}{\arabic{footnote}}
\setcounter{footnote}{0}
\vspace{6ex}

%--------------------<<   abstract   >>--------------------
\begin{abstract}
Non-abelian horizontal symmetry has been considered to
solve potentially SUSY flavor problem, but  simple models are 
suffering from various problems. In this talk, we point out that
(anomalous) $U(1)_A$ gauge symmetry solves all the problems in a 
natural way, especially, in the $E_6$ grand unified theories.
Combining the GUT scenario with anomalous $U(1)_A$ gauge symmetry,
in which doublet-triplet splitting and natural gauge coupling
unification are realized, and
realistic quark and lepton mass matrices are obtained including
bi-large neutrino mixings, complete $E_6\times SU(3)_H$ (or 
$E_6\times SU(2)_H$) GUTs can be obtained, in which 
all the three generation quarks and leptons are
unified into a single multiplet $({\bf 27},{\bf 3})$ (or two multiplets
$({\bf 27},{\bf 2}+{\bf 1})$). This talk is based on Ref.~[1].
\end{abstract}

\end{titlepage}

\section{Problems of Simple Models with Horizontal Symmetry}
First of all, we recall the basic features and the problems
of non-abelian horizontal symmetry,\cite{nonabel,Snonabel} considering 
a simple model with a horizontal symmetry $U(2)_H$, under which
the three generations of quarks and leptons,
$\Psi_i=(\Psi_a,\Psi_3)$ ($a=1,2$) ($\Psi=Q,U,D,L,E,N$), 
transform as ${\bf 2+1}$ and the Higgs fields $H$ and $\bar H$ are 
singlets.
Such a horizontal symmetry is interesting because
only the Yukawa couplings for the third generation are allowed 
by the horizontal symmetry, that accounts for the large top Yukawa coupling,
and because the $U(2)_H$ symmetric interaction
$\int d^4\theta \Psi^{\dagger a}\Psi_a Z^\dagger Z$, where $Z$ 
has a non-vanishing vacuum expectation value (VEV) given by 
$\VEV{Z}\sim \theta^2\tilde m$,
leads to the equal first and second generation sfermion masses,
which may realize suppression of flavor changing neutral currents
(FCNC). 
However, the $U(2)_H$ symmetry must be broken to obtain realistic mass
hierarchical structure of quarks and leptons. 
Therefore, we introduce a doublet Higgs
$\bar F^a$ and an anti-symmetric tensor $A^{ab}$, whose VEVs
$|\VEV{\bar F^a}|=\delta^a_2V$ and $\VEV{A^{ab}}=\epsilon^{ab}v$ 
($\epsilon^{12}=-\epsilon^{21}=1$) break the horizontal symmetry as
\begin{equation}
  U(2)_H \mylimit{V}U(1)_H \mylimit{v} {\rm nothing}.
\end{equation}
The hierarchical structure of the Yukawa couplings is obtained as
\begin{equation}
Y_{u,d,e}\sim\left( \matrix{ 0 & \epsilon' & 0 \cr
               \epsilon' & 0 & \epsilon \cr
               0 & \epsilon & 1}\right),
\end{equation}
where 
$\epsilon\equiv V/\Lambda\gg\epsilon'\equiv v/\Lambda$.
However,  these $U(2)_H$ breaking VEVs lift the degeneracy of
the first and second generation sfermion masses as
\begin{equation}
\tilde m^2_{u,d,e}\sim \tilde m^2\left(\matrix{1 & 0 & 0 \cr
                                      0 & 1+\epsilon^2 & \epsilon \cr
                                      0 & \epsilon & O(1) }\right),
\end{equation}
which are calculated from higher dimensional interactions, like 
$\int d^4\theta(\Psi_a\bar F^a)^\dagger \Psi_b\bar F^bZ^\dagger Z$,
through a non-vanishing VEV $\VEV{\bar F}$. 
These mass matrices lead to the relations
\begin{equation}
\frac{\tilde m_2^2-\tilde m_1^2}{\tilde m^2}
\sim \frac{m_{F2}}{m_{F3}},
\label{Mass}
\end{equation}
where $m_{Fi}$ and $\tilde m_i$ are the masses of the $i$-th generation fermions
and the $i$-th generation sfermions, respectively.
Unfortunately, these relations of this simple model imply
a problematic contribution to the
$\epsilon_K$ parameter in $K$ meson mixing and 
the $\mu\rightarrow e\gamma$ process (problem 1).
Moreover, this simple model gives the similar hierarchical
Yukawa couplings for the up-quark sector, the down-quark sector, 
and the lepton-sector, which are not consistent with experimental results
(problem 2).
In many cases of grand unified theories (GUTs), 
to realize the large neutrino mixing angles 
that have been reported in several recent experiments,\cite{atmos,solar}
the diagonalizing matrices for ${\bf \bar 5}$ fields of $SU(5)$, $V_l$ and
$V_{d_R}$, also have large mixing angles.
In the cases, even if the horizontal symmetry $U(2)_H$ realizes
the degeneracy of the first two generation down squarks such as 
\begin{equation}
\Delta \tilde m_{d_R}^2\equiv \frac{\tilde m_{d_R}^2-\tilde m^2}
{\tilde m^2}\sim \left( \matrix{0&0&0\cr
                                                   0&0&0 \cr
                                                   0&0&\alpha}\right),
\end{equation}
where $\alpha$ is a $O(1)$ parameter, the mixing matrix defined \cite{GGMS}
by
\begin{equation}
\delta_{d_R}\equiv V_{d_R}^\dagger \Delta \tilde m_{d_R}^2V_{d_R},\
\delta_{l_L}\equiv V_{l_L}^\dagger \Delta \tilde m_{l_L}^2V_{l_L}
\end{equation}
have large components $(\delta_{d_R})_{12}$ and $(\delta_{l_L})_{12}$.
It is not so natural to  satisfy the constraints from $\epsilon_K$ in 
$K$ meson mixing
\begin{eqnarray}
\sqrt{|{\rm Im}(\delta_{d_L})_{12}(\delta_{d_R})_{12})|}&\leq&
 2\times 10^{-4}\left(\frac{\tilde m_q}{500 {\rm GeV}}\right) \\
|{\rm Im}(\delta_{d_R})_{12}| & \leq & 1.5\times 10^{-3}
\left(\frac{\tilde m_q}{500 {\rm GeV}}\right)
\label{eK}
\end{eqnarray}
and from $\mu\rightarrow e \gamma$ process
\begin{equation}
|(\delta_{l_L})_{12}|\leq 4\times 10^{-3}\left(
\frac{\tilde m_l}{100 {\rm GeV}}\right)^2,
\label{mu}
\end{equation}
at the weak scale (problem 3), even though 
$(\delta_{d_L})_{12}$ can become fairly small.

In this talk, we show that anomalous $U(1)_A$ gauge 
symmetry,\cite{U(1)} whose anomaly is cancelled by Green-Schwarz 
mechanism,\cite{GS} provides
a natural solution for all these problems.
In a series of papers,\cite{TGUT,BM,MY,Unif} we have emphasized that in solving
various problems in GUT it is important 
that the VEVs
are determined by the anomalous $U(1)_A$ charges as
\begin{equation}
\VEV{O_i}\sim \left\{ 
\begin{array}{ccl}
  \lambda^{-o_i} & \quad & o_i\leq 0 \\
  0              & \quad & o_i>0
\end{array} \right. ,
\label{VEV}
\end{equation}
where the $O_i$ are GUT gauge singlet operators with 
charges
$o_i$, and $\lambda\equiv \VEV{\Theta}/\Lambda\ll 1$.
Here the Froggatt-Nielsen (FN)\cite{FN,abel} field $\Theta$ has 
an anomalous $U(1)_A$ charge of $-1$. 
(In this paper we choose $\Lambda\sim 2\times 10^{16}$ GeV,
 which results
from the natural gauge coupling unification,\cite{Unif}
and $\lambda\sim 0.22$.)
Throughout this paper, we denote all superfields and chiral 
operators by uppercase letters and their anomalous $U(1)_A$ 
charges by the corresponding lowercase letters. When convenient, we
use units in which $\Lambda=1$. 
 Such a vacuum structure is naturally obtained if we introduce generic
interactions even for higher-dimensional operators and if the $F$-flatness 
conditions
determine the scale of the VEVs. And this vacuum structure plays an 
important role in realizing the doublet-triplet splitting\cite{TGUT,MY}
 and natural gauge 
coupling unification\cite{Unif} and 
in avoiding unrealistic GUT relations between 
Yukawa matrices.\cite{TGUT,BM} 
And in this talk, we stress that this vacuum structure
plays an critical role also in solving SUSY flavor problem with horizontal
gauge symmetry.

\section{$SU(5)\times SU(2)_H$}
Let us explain the basic idea of a solution for the problem 1 and 2 
with an $SU(5)$ GUT model with 
$SU(2)_H\times U(1)_A$, though problem 3 still remains in this model.
The field content is given in Table 1.
\begin{center}
Table 1, Typical values of anomalous $U(1)_A$ charges.
The half integer charges play the same role as R-parity.
{\footnotesize
\begin{tabular}{|c|cccccc|cccccc|} 
\hline
   &   $\Psi_a$ & $\Psi_3$ & $T_a$ & $T_3$ & $N_a$ &$ N_3$ & $H$  & $\bar H$ 
   & $F_a$ & $\bar F^a$ & $S$ & $\Theta$ \\
\hline 
 $SU(5)$ & {\bf 10} & {\bf 10} & ${\bf \bar 5}$ & ${\bf \bar 5}$ & {\bf 1}& 
 {\bf 1} & ${\bf 5}$ 
 & ${\bf \bar 5}$ & {\bf 1} & {\bf 1} &{\bf 1}& {\bf 1}  \\
 $SU(2)_H$ & {\bf 2} &{\bf 1} & {\bf 2} &{\bf 1} &{\bf 2} &{\bf 1} &
 ${\bf 1}$ & {\bf 1} &{\bf 2} &${\bf \bar 2}$ &{\bf 1}&{\bf 1}  \\
 $U(1)_A$ & $\frac{13}{2}$ & $\frac{7}{2}$ & $\frac{13}{2}$ & $\frac{11}{2}$ 
 & $\frac{13}{2}$ & $\frac{7}{2}$ & $-7$ & $-7$ & 
 $-2$ & $-3$ & 5 & $-1$  \\
 \hline
\end{tabular}}
\end{center}
The VEV relations (\ref{VEV}) imply $\VEV{\bar FF}\sim \lambda^{-(f+\bar f)}$
which breaks $SU(2)_H$. Actually, the $F$-flatness condition of $S$ with the
superpotential
$
W_S=\lambda^s S(1+\lambda^{f+\bar f}\bar FF),
$
leads to this VEV.
Without loss of generality, we can take
\begin{equation}
|\VEV{\bar F^a}|=|\VEV{F_a}|\sim \delta_{a2}\lambda^{-\frac{1}{2}(f+\bar f)},
\end{equation}
using the $SU(2)_H$ gauge symmetry and its $D$-flatness condition.
Then, because the $SU(2)_H\times U(1)_A$ invariant operators become
\begin{equation}
\lambda^{\psi+\bar f}\Psi_a\VEV{\bar F^a}\sim 
\lambda^{\psi+\Delta f}\Psi_2, \ 
\lambda^{\psi+f}\epsilon^{ab}\Psi_a\VEV{F^b}\sim 
\lambda^{\psi-\Delta f}\Psi_1, \nn
\end{equation}
where $\Delta f\equiv \frac{1}{2}(\bar f-f)$,
it is obvious that with the effective charges defined as
$\tilde x_3\equiv x_3$, $\tilde x_2\equiv x+\Delta f$, 
and $\tilde x_1\equiv x-\Delta f$ for $x=\psi,t,n$,
the Yukawa matrices of the quarks and leptons $Y_{u,d,e,\nu}$ and 
the right-handed neutrino mass matrix $M_{\nu R}$ can be obtained as
\begin{eqnarray}
&&(Y_u)_{ij}\sim \lambda^{\tilde \psi_i+\tilde \psi_j+h},\
(Y_d)_{ij}\sim (Y_e^T)_{ij}\sim \lambda^{\tilde \psi_i+\tilde t_j+\bar h} \\
&&(Y_\nu)_{ij}\sim \lambda^{\tilde t_i+\tilde n_j+h}, \
(M_{\nu R})_{ij}\sim \lambda^{\tilde n_i+\tilde n_j} 
\end{eqnarray}
from the generic 
interactions 
$
W_{\rm fermion}=\tilde \Psi^2\lambda^hH+\tilde \Psi\tilde T
\lambda^{\bar h}\bar H+
\tilde T\tilde N\lambda^hH+\tilde N\tilde N,
$
where 
$\tilde X\equiv \lambda^{x+f}\epsilon^{ab}X_aF_b+\lambda^{x+\bar f}X_a
\bar F^a+\lambda^{x_3}X_3$ for $X=\Psi,T,N$.
Throughout this paper, we omit $O(1)$ coefficients for simplicity.
Then, the  neutrino mass matrix is obtained as
\begin{equation}
(M_{\nu})_{ij}= (Y_\nu)(M_{\nu R})^{-1}(Y_\nu^T)\frac{\VEV{H}^2}{\Lambda}
\sim\lambda^{\tilde t_i+\tilde t_j+2h}\frac{\VEV{H}^2}{\Lambda}.
\end{equation}
Note that the effective charges $\tilde x_1$ can be different from 
$\tilde x_2$, though $x=x_1=x_2$.
When all the Yukawa couplings can be determined by their effective $U(1)_A$
charges, it has been understood not to be difficult to assign their charges 
to obtain realistic quark and lepton mass matrices.\cite{abel,TGUT,BM}
Thus the problem 2 can be solved in this scenario. It is an interesting point
in theories in which Yukawa couplings are determined by  $U(1)$ charges
as in the above,  that the unitary matrices $V_{y_P}$ $(y=u,d,e,\nu$ and 
$P=L,R)$ that diagonalize these 
Yukawa and mass matrices as  
$V_{y_L}^{\dagger}Y_yV_{y_R}=Y_y^{\rm diag}$, the Cabibbo-Kobayashi-Maskawa
matrix $V_{CKM}\equiv V_{d_L}V_{u_L}^\dagger$, and the Maki-Nakagawa-Sakata 
matrix $V_{MNS} \equiv V_{e_L}V_{\nu_L}^\dagger$
are roughly approximated by the matrices 
$(V_{\bf 10})_{ij}\equiv \lambda^{\tilde \psi_i-\tilde \psi_j}$ and
$(V_{\bf \bar 5})_{ij}\equiv \lambda^{\tilde t_i-\tilde t_j}$
as 
$V_{\bf 10}\sim V_{u_L}\sim V_{d_L}\sim V_{u_R} \sim V_{e_R}\sim V_{CKM}$
and
$V_{\bf \bar 5}\sim V_{d_R}\sim V_{e_L}\sim V_{\nu_L}\sim V_{MNS}$.

To see how to solve the problem 1, we examine the sfermion mass-squared 
matrices 
\begin{equation}
\tilde m_y^2=\left(\matrix{\tilde m_{y_L}^2 & A_y^\dagger \cr
                           A_y & \tilde m_{y_R}^2 }\right).
\end{equation}
In this paper, we concentrate on mass mixings through $\tilde m_{y_P}^2$,
because a reasonable assumption, for example SUSY breaking in the hidden 
sector, 
leads to an $A_y$ that is proportional to the Yukawa matrix $Y_y$.\cite{SW}
Roughly speaking, the sfermion mass squared matrix is given by
$\tilde m_{y_P}^2\sim \tilde m^2 {\rm diag} (1,1,O(1))$, and 
the correction 
$\Delta_{y_P}\equiv(\tilde m_{y_P}^2-\tilde m^2)/(\tilde m_{y_P})^2$ 
in the model described by Table I is approximately given by 
\begin{equation}
\Delta_{\bf 10}=\left(
\matrix{\lambda^5 & \lambda^6 & \lambda^{3.5}\cr
 \lambda^6 &\lambda^5 &  \lambda^{2.5} \cr
 \lambda^{3.5} & \lambda^{2.5}& R_{\bf 10} \cr }
\right),
\Delta_{\bf \bar 5}=\left(
\matrix{\lambda^5 & \lambda^6 & \lambda^{3.5} \cr
 \lambda^6 & \lambda^5 & \lambda^{4.5} \cr
 \lambda^{3.5} &  \lambda^{4.5}& R_{\bf \bar 5} \cr }
\right)
\end{equation}
for {\bf 10} fields and ${\bf \bar 5}$ fields.
Here $R_{\bf 10, \bar 5}\sim O(1)$.
For example, $(\Delta_{\bf \bar 5})_{12}$ can be derived using
the interaction 
$\int d^4\theta \lambda^{|f-\bar f|}(T\bar F)^\dagger (TF)Z^\dagger Z$.
Note that 
$(\tilde m_{d2}^2-\tilde m_{d1}^2)/\tilde m_d^2\sim 
(m_s/m_b)^2$.
The essential point here is that the hierarchy originated from the VEVs
$|\VEV{F}|=|\VEV{\bar F}|\sim \lambda^{-\frac{1}{2}(f+\bar f)}$
is almost cancelled by the enhancement factors $\lambda^f$ and
$\lambda^{\bar f}$ in the superpotential, but not in the K\"ahler potential.
Thus, the Yukawa hierarchy in the superpotential becomes milder, that
improves the unrealistic relations (\ref{Mass}).

Unfortunately, as discussed in the previous section,
because the neutrino mixing angles 
are large and because $R_{\bar 5}\sim O(1)$,
the suppression of FCNC processes may not be sufficient (problem 3).
The various FCNC processes constrain the mixing matrices defined by
$\delta_{y_P}\equiv V_{y_P}^\dagger \Delta_{y_P}V_{y_P}$.\cite{GGMS}
In the model in Table 1, the
mixing matrices are approximated as  
\begin{equation}
\delta_{\bf 10}=\left(\matrix{\lambda^5 & \lambda^6 & \lambda^{3.5} \cr
                          \lambda^6 & \lambda^5 & \lambda^{2.5} \cr
                          \lambda^{3.5} & \lambda^{2.5} & R_{\bf 10} }
             \right), \ 
\delta_{\bf \bar 5}=R_{\bf \bar 5}
          \left(\matrix{\lambda^3 & \lambda^2 & \lambda^{1.5} \cr
                          \lambda^2 & \lambda & \lambda^{0.5} \cr
                          \lambda^{1.5} & \lambda^{0.5} & 1 }
             \right)
\label{delta}
\end{equation}
at the GUT scale. The constraints at the weak scale from 
$\epsilon_K$ in $K$ meson mixing (\ref{eK})
requires scalar quark masses larger than
1 TeV, because in this model
$\sqrt{|(\delta_{d_L})_{12}(\delta_{d_R})_{12})|}\sim \lambda^4(\eta_q)^{-1}$
and $|(\delta_{d_R})_{12}|\sim \lambda^2(\eta_q)^{-1}$, where
we take a renormalization factor $\eta_q\sim 6$.\footnote{
The renormalization factor is strongly dependent on the ratio of the gaugino 
mass to the scalar fermion mass and the model below the GUT scale.
If the model is MSSM and the ratio at the GUT scale is 1, 
then $\eta_q=6\sim 7$.}
And the constraint from the $\mu\rightarrow e\gamma$ process (\ref{mu})
requires scalar lepton masses larger than 300 GeV,
because 
$|(\delta_{l_L})_{12}|\sim \lambda^2$ in this model.

\section{$E_6\times SU(2)_H$}
It is obvious that if all the three generation  ${\bf \bar 5}$ scalar
fermions have degenerate masses, the problem 3 can be solved. 
In this section, we show that in $E_6$ GUT, such scalar fermion mass
structure can be obtained in a natural way.
First, note that under $E_6\supset SO(10)\supset SU(5)$, 
the fundamental representation
${\bf 27}$ is divided as
\begin{equation}
{\bf27} \rightarrow {\bf 16}[ {\bf 10} +{\bf \bar 5}
+{\bf 1}]
+{\bf 10}[{\bf \bar 5'}+{\bf 5}]
+{\bf 1}[{\bf 1}].
\end{equation}
To break $E_6$ into $SU(5)$, two pairs of ${\bf 27}$ and ${\bf \overline{27}}$
are introduced. (And an adjoint Higgs $A({\bf 78})$ is needed to break $SU(5)$
into the standard model gauge group, but here the Higgs does not play an 
important role and we do not address the Higgs.) The VEVs 
$|\VEV{\Phi}|=|\VEV{\bar \Phi}|\sim \lambda^{-\frac{1}{2}(\phi+\bar \phi)}$
 break $E_6$
into $SO(10)$, which is broken into $SU(5)$ by the VEVs 
$|\VEV{C}|=|\VEV{\bar C}|\sim \lambda^{-\frac{1}{2}(c+\bar c)}$.
And as matter fields, three fundamental representation fields 
$\Psi_i({\bf 27})$ ($i=1,2,3$) are introduced, which include
 $3\times {\bf 5}+6\times {\bf \bar 5}$ of $SU(5)$.
Note that only three of the six ${\bf \bar 5}$ become massless, which
are determined by the $3\times 6$ mass matrix obtained from the 
interactions 
$W=\lambda^{\psi_i+\psi_j+\phi}\Psi_i\Psi_j\Phi
+\lambda^{\psi_i+\psi_j+c}\Psi_i\Psi_jC$. 
It is essential that  because the third generation fields
have larger Yukawa 
couplings than the first and second generation fields 
($\psi_3<\psi_1,\psi_2$),
the third generation fields ${\bf \bar 5}_3$ and ${\bf \bar 5'}_3$
have larger masses in the $3\times 6$ mass matrix than
the first and second generation fields ${\bf \bar 5}_a$ and ${\bf \bar 5'}_a$
($a=1,2$), respectively.
Therefore,
it is natural that
these three massless ${\bf \bar 5}$ fields come from the first and second 
generation fields, $\Psi_1$ and $\Psi_2$, 
as discussed in Ref.~[12]. %\cite{BM}.
If the first two multiplets become the doublet $\Psi({\bf 27}, {\bf 2})$ under
$SU(2)_H$ in this $E_6$ GUT, 
then it is obvious that the sfermion masses for these three massless modes
 ${\bf \bar 5}$ are equal at leading order, because the massless modes
 are originated from
 a single multiplet $({\bf 27},{\bf 2})$.
This sfermion mass structure is nothing but what is required to solve
the problem 3.

Of course, the breakings of the $E_6$ and the horizontal symmetry $SU(2)_H$
lift the degeneracy. To estimate the corrections, we fix a model.
If we adopt the anomalous $U(1)_A$ charges as $(f,\bar f)=(-2,-3)$ and
$(\psi,\psi_3,\phi,\bar \phi,c,\bar c)=(5,2,-4,2,-5,-2)$ (noting
that odd R-parity
is required for the matter fields $\Psi$ and $\Psi_3$), 
the massless modes become 
$({\bf \bar 5_1}, {\bf \bar 5_2}, {\bf \bar 5'_1}
+\lambda^\Delta{\bf \bar 5_3})$, where
$\Delta=\tilde \psi_1-\tilde \psi_3+\frac{1}{2}(\phi-\bar \phi-c+\bar c)=2$.
As discussed in Ref.~[13], %\cite{MY}, 
the massless mode $ {\bf \bar 5'_1}+\lambda^\Delta{\bf \bar 5_3}$ has
Yukawa couplings only through the mixing with ${\bf \bar 5_3}$,
because the ${\bf \bar 5'}$ fields have no direct Yukawa couplings 
with the Higgs fields $H$ and $\bar H$, which are included in
${\bf 10_{\Phi}}$ of $SO(10)$ in many cases. 
Then the structure of the quark and lepton Yukawa matrices
becomes the same as that found in the previous $SU(5)$ model.
As discussed above, all the sfermion masses for ${\bf \bar 5}$ become
equal at the leading order in this model.
The correction to the sfermion masses  $\delta \tilde m_{\bf \bar 5}$ 
can be approximated from the
higher dimensional interactions as
\begin{equation}
\frac{\delta \tilde m_{\bf \bar 5}^2}{\tilde m^2}
\sim \left(
\matrix{\lambda^5 & \lambda^6 & \lambda^{5.5} \cr
 \lambda^6 & \lambda^5 &  \lambda^{4.5} \cr
 \lambda^{5.5} & \lambda^{4.5}& \lambda^2  }
\right),
\end{equation}
which leads to the same $\delta_{\bf \bar 5}$ as that in Eq. (\ref{delta})
if we take $R_{\bf\bar 5}=\lambda^2$. This decreases the lower limit
of the scalar quark mass to an acceptable level, 250 GeV. 
Note that $R_{\bar 5}$ is determined by the $E_6$ breaking scale,
$\VEV{\bar \Phi\Phi}\sim \lambda^{-(\phi+\bar \phi)}$, because 
the $E_6$ breaking VEVs lift the degeneracy through the interaction
$\int d^4\theta \Psi^\dagger \Phi^\dagger \Psi\Phi Z^\dagger Z$.
In the model in Table 1, 
$R_{\bf \bar 5}=(\delta \tilde m_{\bf \bar 5}^2/\tilde m^2)_{33}\sim 
\VEV{\bar \Phi\Phi}\sim \lambda^{-(\phi+\bar \phi)}\sim \lambda^2$, but
we can make various models with different $R_{\bf \bar 5}$ by taking
various charges $\phi+\bar \phi$.

It is surprising that in $E_6$ GUT with anomalous $U(1)_A$ symmetry
the horizontal gauge group can be extended into $SU(3)_H$.
In the models, the three generations of quarks and leptons
can be unified into a single multiplet, $\Psi({\bf 27},{\bf 3})$.
Supposing that the horizontal gauge symmetry $SU(3)_H$ is broken 
by the VEVs of two pairs of Higgs fields $F_i({\bf 1},{\bf 3})$ and 
$\bar F_i({\bf 1},{\bf \bar 3})$ $(i=2,3)$ as
\begin{equation}
|\VEV{F_{ia}}|=|\VEV{\bar F_i^a}|\sim 
\delta_i^a\lambda^{-\frac{1}{2}(f_i+\bar f_i)},
\end{equation}
the effective charges can be defined from the relations 
\begin{eqnarray}
&&\lambda^{\psi+\bar f_i}\Psi_a\VEV{\bar F_i^a}\sim 
\lambda^{\psi+\frac{1}{2}(\bar f_i-f_i)}\Psi_i (i=2,3), \\
&&\lambda^{\psi+f_2+f_3}\epsilon^{abc}\Psi_a\VEV{F_{2b}F_{3c}}\sim 
\lambda^{\psi-\frac{1}{2}(\bar f_2-f_2+\bar f_3-f_3)}\Psi_1, \nn
\end{eqnarray}
as
\begin{equation}
\tilde \psi_i\equiv \psi+\frac{1}{2}(\bar f_i-f_i),\quad
\tilde \psi_1\equiv \psi-\frac{1}{2}(\bar f_2-f_2+\bar f_3-f_3).
\end{equation}
Note that in order to realize $O(1)$ top Yukawa coupling, $SU(3)_H$
must be broken at the cutoff scale, namely, $f_3+\bar f_3=0$.
To obtain the same mass matrices of quarks and leptons as in the previous
$E_6\times SU(2)_H$ model, the effective charges must be taken as
$(\tilde \psi_1,\tilde \psi_2,\tilde \psi_1)=(11/2,9/2,2)$.
For example, a set of charges $(f_3,\bar f_3,f_2,\bar f_2)=(2,-2,-3,-2)$ 
and $\psi=4$ is satisfied with the above conditions.
[The model obtained by choosing  
$(f_3,\bar f_3,f_2,\bar f_2)=(2,-3,-4,-3)$, $\psi=13/2$, and 
$(\phi,\bar \phi,c,\bar c)=(-7,3,-8,0)$ may be more interesting, because
mass matrices for quarks and leptons that are essentially the same
as those in Ref.~[12] %\cite{BM}
are obtained if we set $\lambda^{1.5}=0.22$. ]

For both models $E_6\times SU(2)_H$ and $E_6\times SU(3)_H$, if we add
a Higgs sector that breaks $E_6$ into the gauge group of the standard model,
as in Ref.~[13], %\cite{MY},  
then we can obtain complete
$E_6\times SU(2)_H$ and $E_6\times SU(3)_H$ GUT, in which 
the degeneracy of the sfermion masses for ${\bf\bar 5}$ fields
is naturally obtained.
As discussed in Refs.~[12]$-$[14], %\cite{BM,MY}, 
these models yield not only 
realistic quark and
lepton mass matrices but also doublet-triplet splitting and 
natural gauge coupling unification.

Because the $SU(3)_H$ symmetry must be broken at the cutoff scale 
to realize $O(1)$ top Yukawa coupling, the degeneracy of sfermion
masses between the third generation fields $\Psi_3$ and the first
and second generation fields $\Psi_a$ $(a=1,2)$ is not guaranteed.
Therefore, the $E_6\times SU(3)_H$ GUT gives the same predictions
for the structure of sfermion masses as $E_6\times SU(2)_H$ GUT. 
Roughly speaking,
 all the sfermion fields have nearly equal masses, except the third generation
 fields included in {\bf 10} of $SU(5)$.
It must be an interesting subject to study the predictions on 
FCNC processes (for example, $B$-physics\cite{Okada}) from such a 
special structure of sfermion masses.
More precisely, this degeneracy is lifted by $D$-term 
contributions of $SU(3)_H$ and $E_6$. 
Though the contributions are strongly dependent on the concrete
models for SUSY breaking and on GUT models and 
and some of them must be small in order to suppress the 
FCNC processes, it is important
to test these GUT models
with precisely measured masses of sfermions, as discussed 
in Ref.~[17].%\cite{KMY}.

  N.M. is supported in part by Grants-in-Aid for Scientific 
Research from the Ministry of Education, Culture, Sports, Science 
and Technology of Japan.

\end{document}